\documentclass[runningheads]{llncs}
\usepackage[T1]{fontenc}
\usepackage{amsmath,amssymb}
\usepackage{booktabs,array,graphicx}
\usepackage{tikz}
\usepackage{colortbl}
\usetikzlibrary{arrows.meta,positioning,fit,backgrounds,calc,shapes.geometric,decorations.pathreplacing}
\usepackage{enumitem}
\usepackage{hyperref}
\hypersetup{hidelinks}

\tikzset{
  fbox/.style={draw, rounded corners=2pt, align=center, inner sep=4pt,
               font=\small, minimum height=7mm},
  fdark/.style={fbox, fill=black!7},
  fmid/.style={fbox, fill=black!14},
  farrow/.style={-{Latex[length=2mm]}, thick},
  fdash/.style={draw, dashed, rounded corners=3pt, inner sep=6pt},
  flab/.style={font=\footnotesize\itshape, align=center},
}

\definecolor{ctask}{HTML}{475569}\definecolor{ctaskf}{HTML}{F1F5F9}
\definecolor{csolver}{HTML}{0369A1}\definecolor{csolverf}{HTML}{E0F2FE}
\definecolor{cenv}{HTML}{047857}\definecolor{cenvf}{HTML}{D1FAE5}
\definecolor{cver}{HTML}{4338CA}\definecolor{cverf}{HTML}{E0E7FF}
\definecolor{cgt}{HTML}{B45309}\definecolor{cgtf}{HTML}{FEF3C7}
\definecolor{cgen}{HTML}{7C2D12}\definecolor{cgenf}{HTML}{FFF7ED}
\definecolor{creward}{HTML}{BE123C}
\tikzset{
  ar/.style={-{Latex[length=2.2mm]}, line width=0.9pt, black!75},
  dar/.style={-{Latex[length=2mm]}, line width=0.8pt, dashed, black!50},
  lab/.style={font=\scriptsize, align=center, inner sep=1.5pt, fill=white},
  hd/.style={font=\small\bfseries, align=center},
  sub/.style={font=\scriptsize, align=center, black!65},
  num/.style={circle, fill=black!80, text=white, font=\scriptsize\bfseries, inner sep=0.6pt, minimum size=3.6mm},
}
\newcommand{\chain}[4]{\foreach \i in {1,...,#3} {
  \fill[white, draw=#4, rounded corners=1pt, line width=0.6pt] (#1+\i*0.5-0.5,#2-0.17) rectangle ++(0.36,0.34);
  \draw[#4!60] (#1+\i*0.5-0.43,#2-0.06) -- ++(0.22,0) (#1+\i*0.5-0.43,#2+0.04) -- ++(0.14,0);
  \ifnum\i<#3 \draw[#4, line width=0.9pt] (#1+\i*0.5-0.14,#2) -- ++(0.14,0); \fi}}
\newcommand{\gtgraph}{
  \begin{scope}[shift={(-0.8,0.02)}, every node/.style={draw=cgt!80!black, fill=white, inner sep=0pt, minimum size=1.9mm, circle}]
   \node (k0) at (0,0) {}; \node (k1) at (0.45,0.28) {}; \node (k2) at (0.45,-0.28) {}; \node (k3) at (0.9,-0.28) {}; \node (k4) at (0.9,0.28) {}; \node (k5) at (1.35,0) {};
   \foreach \a/\b in {k0/k1,k0/k2,k2/k3,k1/k4,k4/k5,k3/k5} {\draw[cgt!80!black, -{Latex[length=0.8mm]}] (\a)--(\b);}
  \end{scope}
  \draw[cgt!80!black, line width=1pt] (1.2,0.36) circle (0.16); \draw[cgt!80!black, line width=1pt] (1.12,0.36) -- (1.18,0.3) -- (1.29,0.44);}
\newcommand{\templates}{\fill[white, draw=cgen, line width=0.6pt] (-0.25,-0.2) rectangle (0.25,0.2);
  \fill[white, draw=cgen, line width=0.6pt] (-0.35,-0.1) rectangle (0.15,0.3);
  \fill[white, draw=cgen, line width=0.6pt] (-0.45,0.0) rectangle (0.05,0.4);
  \draw[cgen!60] (-0.38,0.28) -- (-0.05,0.28) (-0.38,0.18) -- (-0.12,0.18);}
\newcommand{\listicon}{\fill[white, draw=cgen, line width=0.6pt, rounded corners=1pt] (-0.32,-0.32) rectangle (0.32,0.32);
  \foreach \yy in {0.16,0.0,-0.16} {\fill[cgen!70] (-0.22,\yy) circle (0.03); \draw[cgen!70] (-0.14,\yy) -- (0.22,\yy);}}
\newcommand{\scrollicon}{\fill[white, draw=cgen, line width=0.6pt, rounded corners=1pt] (-0.3,-0.32) rectangle (0.3,0.32);
  \draw[cgen!70] (-0.2,0.15) -- (0.2,0.15) (-0.2,0.0) -- (0.15,0.0) (-0.2,-0.15) -- (0.1,-0.15);}
\newcommand{\person}[1]{\fill[#1!15, draw=#1, line width=0.8pt] (0,0.32) circle (0.19);
  \fill[#1!15, draw=#1, line width=0.8pt] (-0.36,-0.32) .. controls (-0.36,0.05) and (0.36,0.05) .. (0.36,-0.32) -- cycle;}

\newcommand{\resAbfPlainQuarterCell}{\textbf{39}}
\newcommand{\resCodexPlainQuarterCell}{25}
\newcommand{\resClaudePlainQuarterCell}{9}
\newcommand{\resBfsPlainQuarterCell}{31}
\newcommand{\resDfsPlainQuarterCell}{31}
\newcommand{\resGreedyPlainQuarterCell}{26}
\newcommand{\resAbfPlainHalfCell}{\textbf{70}}
\newcommand{\resCodexPlainHalfCell}{41}
\newcommand{\resClaudePlainHalfCell}{11}
\newcommand{\resBfsPlainHalfCell}{64}
\newcommand{\resDfsPlainHalfCell}{60}
\newcommand{\resGreedyPlainHalfCell}{46}
\newcommand{\resAbfPlainFullCell}{75}
\newcommand{\resCodexPlainFullCell}{44}
\newcommand{\resClaudePlainFullCell}{11}
\newcommand{\resBfsPlainFullCell}{\textbf{91}}
\newcommand{\resDfsPlainFullCell}{79}
\newcommand{\resGreedyPlainFullCell}{58}
\newcommand{\resAbfOpsClaimedCell}{83}
\newcommand{\resCodexOpsClaimedCell}{61}
\newcommand{\resClaudeOpsClaimedCell}{78}
\newcommand{\resBfsOpsClaimedCell}{85}
\newcommand{\resDfsOpsClaimedCell}{88}
\newcommand{\resGreedyOpsClaimedCell}{\textbf{100}}
\newcommand{\resFidMinPct}{94}
\newcommand{\resSuite}{31}

\newcommand{\resBfsRecall}{0.88}
\newcommand{\resBfsRecallFin}{0.91}

\newcommand{\resBfsRepDef}{0.7}
\newcommand{\resBfsOpsClaimed}{85}

\newcommand{\resBfsTpArr}{87}

\newcommand{\resBfsTpPast}{367}

\newcommand{\resDfsRepDef}{1.1}

\newcommand{\resGreedyRepDef}{0.0}

\newcommand{\resAbfRecall}{0.75}

\newcommand{\resAbfFid}{0.98}
\newcommand{\resAbfFidStrict}{0.98}

\newcommand{\resAbfShare}{47}
\newcommand{\resAbfUnused}{53}

\newcommand{\resAbfSmall}{0.82}

\newcommand{\resAbfLarge}{0.69}

\newcommand{\resAbfPairN}{23}

\newcommand{\resAbfPairSD}{0.18}

\newcommand{\resAbfBeatsBfs}{7}

\newcommand{\resAbfPlainQuarter}{39}

\newcommand{\resAbfAheadOwnStop}{26}

\newcommand{\resAbfEnvCross}{54}
\newcommand{\resAbfRhoOutdeg}{-0.66}
\newcommand{\resAbfRhoEdges}{-0.41}
\newcommand{\resAbfRepDef}{2.5}
\newcommand{\resAbfOpsClaimed}{83}

\newcommand{\resAbfTpRefusalPct}{89}

\newcommand{\resAbfTpArr}{63}

\newcommand{\resCodexRecall}{0.47}

\newcommand{\resCodexPairN}{31}

\newcommand{\resCodexPairSD}{0.17}

\newcommand{\resCodexRepDef}{1.0}

\newcommand{\resCodexClaimedCrossing}{44}

\newcommand{\resClaudeRecall}{0.10}

\newcommand{\resClaudeRecallLenient}{0.54}

\newcommand{\resClaudeFidStrict}{0.23}

\newcommand{\resClaudeRepDef}{1.8}

\newcommand{\resTracerTpRefusalPctMax}{0}
\newcommand{\resEnvQuarter}{35}

\newcommand{\suiteHiddenArrivals}{113}

\newcommand{\suiteMaxOutDegMax}{28}

\newcommand{\suiteRosterDisclosed}{20}

\newcommand{\resultpercent}[1]{%
  \begingroup
  \pgfmathparse{100*(#1)}%
  \pgfmathprintnumber[fixed,precision=0]{\pgfmathresult}\%\endgroup}

\title{ABF-Bench: A Verifiable Benchmark for\\Agentic Blockchain Forensics}
\titlerunning{ABF-Bench}
\author{Xiaohang Yu\inst{1} \and Hejia Geng\inst{2} \and William Knottenbelt\inst{1}}
\authorrunning{X. Yu et al.}
\institute{Imperial College London\\
\email{\{x.yu21,w.knottenbelt\}@imperial.ac.uk} \and
University of Oxford}

\begin{document}
\maketitle
\setcounter{footnote}{0}
\begin{abstract}
Blockchain forensics is adversarial and open-ended: money launderers
split, swap, and bridge funds, cut their trail at mixers or custodial
services, and change tactics from one incident to the next.
Investigators must keep adapting their tracing in response. Yet the
immutability of public ledgers preserves every recorded movement in
public view, making much of the investigation independently verifiable.
We introduce ABF-Bench, an interactive benchmark that makes
evidence-backed fund tracing a verifiable agent task, in which every
claimed movement must cite ledger records the agent retrieved. A
case-to-task pipeline builds well-posed tasks across ledgers:
conservative rules halt tracing wherever the records leave a
continuation unresolved, and each task's reference graph is the answer
those rules determine. Built from
real laundering activity, the tasks combine multi-hop transfers, DeFi
exchanges, cross-chain round trips, custodial and mixer deposits, and
address poisoning. We also propose the ABF agent harness, built around
a working memory, and evaluate it alongside general-purpose agents and
fixed-order tracers. The harness reaches
\resultpercent{\resAbfRecall} edge recall, against
\resultpercent{\resCodexRecall} for Codex with the same model and
\resultpercent{\resClaudeRecall} for Claude Code, which loses recall to
overconfidence: it cites records it left unretrieved. Compared with fixed-order tracers, the harness
acquires evidence faster than the best of them early in the budget and
recognizes undisclosed custodial services that they walk past
(\resAbfTpRefusalPct\% vs.\ \resTracerTpRefusalPctMax\%),
with \resultpercent{\resAbfFid} fidelity. However, agents tend to stop
early and lose recall on larger tasks, which holds their final recall
below the best tracer. Beyond exposing these gaps, ABF-Bench opens
blockchain forensics to LLM agents as a verifiable domain, where the
verdicts that score agents also offer potential training signals.
\end{abstract}

\section{Introduction}
\label{sec:intro}

Fund tracing determines where stolen value came from and where it
went, and it is central to anti-money-laundering compliance and
incident response on blockchains. In 2024 alone, attackers stole \$2.2
billion from cryptocurrency platforms~\cite{chainalysis2024}.
Launderers adapt their methods to defeat tracing, so fixed
procedures fall behind new techniques (Section~\ref{sec:tests}). An
investigation can cross assets, contract routers, and ledgers before
reaching a service whose internal records are unavailable, and the
investigator must decide which records to inspect and which
conclusions they support. LLM agents can make these
decisions, but forensic conclusions must remain checkable by another
investigator.

Public ledgers turn part of this work into a verifiable reasoning task:
given the relevant records and operation rules, a verifier can assess
a claimed movement without trusting the investigator's reasoning.
Locating those records still requires well-chosen queries,
interpretation across protocols, and attention to unfinished leads.

Verifiable tasks of this kind have driven recent agent progress.
Reinforcement learning from verifiable rewards underlies the
post-training of current reasoning models~\cite{deepseekr1} and
LLM agents~\cite{zhang2026landscape}. Benchmark tasks with mechanical
checks, led by SWE-bench's
test-verified code fixes~\cite{swebench}, double as training
environments.
Frontier models now approach ceiling scores on many of these suites,
and successors such as Humanity's Last Exam and FrontierMath
commission expert problems to stay ahead of
them~\cite{hle,frontiermath}. Fund tracing adds a domain where
difficulty needs no curation: real adversaries produce the instances,
and the ledger keeps every claim checkable. A benchmark here therefore
serves both fields: it advances LLM agents broadly and strengthens
agentic investigation in forensics itself.

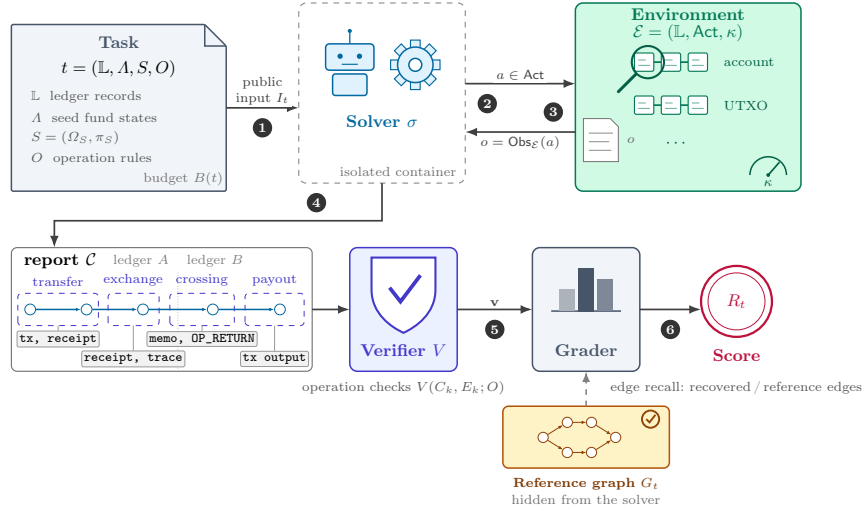
\begin{figure}[t]
\centering
\resizebox{0.93\linewidth}{!}{%
\begin{tikzpicture}[font=\small]

% ================= TASK
\begin{scope}[shift={(0,0)}]
 \fill[ctaskf, draw=ctask, line width=0.8pt, rounded corners=1pt] (-2.0,-1.55) -- (-2.0,1.55) -- (1.6,1.55) -- (2.0,1.15) -- (2.0,-1.55) -- cycle;
 \draw[ctask, line width=0.8pt] (1.6,1.55) -- (1.6,1.15) -- (2.0,1.15);
 \node[hd, ctask] at (0,1.22) {Task};
 \node[font=\small] at (0,0.72) {$t=(\mathbb{L},\Lambda,S,O)$};
 \node[sub, anchor=west] at (-1.75,0.2)   {$\mathbb{L}$\ \ ledger records};
 \node[sub, anchor=west] at (-1.75,-0.18) {$\Lambda$\ \ seed fund states};
 \node[sub, anchor=west] at (-1.75,-0.56) {$S=(\Omega_S,\pi_S)$};
 \node[sub, anchor=west] at (-1.75,-0.94) {$O$\ \ operation rules};
 \node[font=\scriptsize, black!55] at (1.2,-1.32) {budget $B(t)$};
\end{scope}

% ================= SOLVER
\begin{scope}[shift={(4.9,0.15)}]
 \draw[black!45, dashed, rounded corners=4pt, line width=0.6pt] (-1.55,-1.55) rectangle (1.55,1.8);
 \node[font=\scriptsize, black!55, anchor=south east] at (1.5,-1.53) {isolated container};
 \begin{scope}[shift={(-0.6,0.6)}]
  \draw[csolver, line width=0.7pt] (0,0.44) -- (0,0.64); \fill[csolver] (0,0.68) circle (1.1pt);
  \fill[csolverf, draw=csolver, line width=0.8pt, rounded corners=2pt] (-0.45,-0.05) rectangle (0.45,0.44);
  \fill[csolver] (-0.18,0.21) circle (1.4pt); \fill[csolver] (0.18,0.21) circle (1.4pt);
  \draw[csolver, line width=0.6pt] (-0.15,0.06) -- (0.15,0.06);
  \fill[csolverf, draw=csolver, line width=0.8pt, rounded corners=1.5pt] (-0.32,-0.52) rectangle (0.32,-0.12);
 \end{scope}
 \begin{scope}[shift={(0.62,0.65)}]
  \foreach \a in {0,45,...,315} {\fill[csolverf, draw=csolver, line width=0.6pt, rotate=\a] (-0.09,0.29) rectangle (0.09,0.46);}
  \fill[csolverf, draw=csolver, line width=0.8pt] (0,0) circle (0.33);
  \fill[white, draw=csolver, line width=0.8pt] (0,0) circle (0.12);
 \end{scope}
 \node[hd, csolver] at (0,-0.4) {Solver $\sigma$};
\end{scope}
\coordinate (Aw) at (3.35,0.0); \coordinate (Ae) at (6.45,0.0); \coordinate (As) at (4.9,-1.4);

% ================= ENVIRONMENT
\begin{scope}[shift={(10.6,0.15)}]
 \fill[cenvf, draw=cenv, line width=0.8pt, rounded corners=4pt] (-2.1,-1.7) rectangle (2.1,1.8);
 \node[hd, cenv, font=\footnotesize\bfseries, align=center] at (0,1.4) {Environment\\[-2pt]{\mdseries $\mathcal{E}=(\mathbb{L},\mathsf{Act},\kappa)$}};
 \chain{-1.0}{0.7}{3}{cenv} \node[font=\scriptsize, cenv!80!black, anchor=west] at (0.55,0.7) {account};
 \chain{-1.0}{-0.1}{3}{cenv} \node[font=\scriptsize, cenv!80!black, anchor=west] at (0.55,-0.1) {UTXO};
 \node[font=\small, cenv!70!black] at (-0.2,-0.8) {$\cdots$};
 \draw[cenv!80!black, line width=1.1pt] (-0.75,0.7) circle (0.3);
 \draw[cenv!80!black, line width=1.6pt, line cap=round] (-0.97,0.48) -- (-1.3,0.15);
 \fill[white, draw=black!55, line width=0.5pt] (-1.95,-1.15) -- (-1.95,-0.35) -- (-1.45,-0.35) -- (-1.3,-0.5) -- (-1.3,-1.15) -- cycle;
 \draw[black!45] (-1.85,-0.62) -- (-1.4,-0.62) (-1.85,-0.78) -- (-1.47,-0.78) (-1.85,-0.94) -- (-1.43,-0.94);
 \node[font=\scriptsize, black!60, anchor=west] at (-1.25,-0.75) {$o$};
 \draw[cenv!80!black, line width=0.8pt] (1.15,-1.4) arc (180:0:0.34);
 \draw[cenv!80!black, line width=0.9pt] (1.49,-1.4) -- (1.7,-1.16);
 \fill[cenv!80!black] (1.49,-1.4) circle (1pt);
 \node[font=\scriptsize, cenv!80!black] at (1.49,-1.56) {$\kappa$};
\end{scope}
\coordinate (Ew) at (8.5,0.0);

% ================= REPORT (row 2, left; the solver's output, an artifact)
\begin{scope}[shift={(-2.0,-3.75)}]
 \fill[white, draw=black!55, rounded corners=3pt, line width=0.6pt] (0,-1.15) rectangle (5.6,1.15);
 \node[font=\small\bfseries, anchor=west] at (0.12,0.9) {report $\mathcal{C}$};
 \draw[black!30, densely dotted] (3.1,-1.1) -- (3.1,0.72);
 \node[font=\scriptsize, black!50, anchor=east] at (3.05,0.9) {ledger $A$};
 \node[font=\scriptsize, black!50, anchor=west] at (3.15,0.9) {ledger $B$};
 \begin{scope}[every node/.style={draw=csolver, fill=white, inner sep=0pt, minimum size=2mm, circle}]
  \node (m0) at (0.35,0.0) {}; \node (m1) at (1.4,0.0) {}; \node (m2) at (2.45,0.0) {}; \node (m3) at (3.75,0.0) {}; \node (m4) at (5.0,0.0) {};
 \end{scope}
 \foreach \a/\b in {m0/m1,m1/m2,m2/m3,m3/m4} {\draw[csolver, -{Latex[length=0.9mm]}, thick] (\a)--(\b);}
 \draw[cver, dashed, rounded corners=2pt] (0.12,-0.3) rectangle (1.62,0.3);
 \draw[cver, dashed, rounded corners=2pt] (1.8,-0.3) rectangle (2.75,0.3);
 \draw[cver, dashed, rounded corners=2pt] (2.95,-0.3) rectangle (4.15,0.3);
 \draw[cver, dashed, rounded corners=2pt] (4.35,-0.3) rectangle (5.45,0.3);
 \node[font=\scriptsize, cver, anchor=south] at (0.87,0.32) {transfer};
 \node[font=\scriptsize, cver, anchor=south] at (2.27,0.32) {exchange};
 \node[font=\scriptsize, cver, anchor=south] at (3.55,0.32) {crossing};
 \node[font=\scriptsize, cver, anchor=south] at (4.9,0.32) {payout};
 \node[font=\scriptsize\ttfamily, draw=black!40, fill=black!5, inner sep=1.3pt, rounded corners=1pt] (c1) at (0.87,-0.55) {tx, receipt};
 \node[font=\scriptsize\ttfamily, draw=black!40, fill=black!5, inner sep=1.3pt, rounded corners=1pt] (c2) at (2.27,-0.92) {receipt, trace};
 \node[font=\scriptsize\ttfamily, draw=black!40, fill=black!5, inner sep=1.3pt, rounded corners=1pt] (c3) at (3.55,-0.55) {memo, OP\_RETURN};
 \node[font=\scriptsize\ttfamily, draw=black!40, fill=black!5, inner sep=1.3pt, rounded corners=1pt] (c4) at (4.9,-0.92) {tx output};
 \draw[black!40] (c1.north) -- (0.87,-0.3); \draw[black!40] (c2.north) -- (2.27,-0.3);
 \draw[black!40] (c3.north) -- (3.55,-0.3); \draw[black!40] (c4.north) -- (4.9,-0.3);
\end{scope}
\coordinate (Cn) at (-1.2,-2.6); \coordinate (Ce) at (3.6,-3.75);

% ================= VERIFIER (row 2; an actor box like the solver and the environment)
\begin{scope}[shift={(5.3,-3.75)}]
 \fill[cverf!60, draw=cver, line width=0.8pt, rounded corners=4pt] (-1.0,-1.15) rectangle (1.0,1.15);
 \begin{scope}[shift={(0,0.3)}, scale=0.7]
  \fill[white, draw=cver, line width=1.1pt] (-0.85,0.95) -- (0.85,0.95) -- (0.85,0.1) .. controls (0.85,-0.55) and (0.32,-0.9) .. (0,-1.05) .. controls (-0.32,-0.9) and (-0.85,-0.55) .. (-0.85,0.1) -- cycle;
  \draw[cver, line width=2pt] (-0.37,0.15) -- (-0.08,-0.22) -- (0.45,0.48);
 \end{scope}
 \node[hd, cver] at (0,-0.75) {Verifier $V$};
 \node[sub] at (0,-1.45) {operation checks $V(C_k,E_k;O)$};
\end{scope}
\coordinate (Vw) at (4.3,-3.75); \coordinate (Ve) at (6.3,-3.75);

% ================= GRADER (row 2; scores the verdicts against the ground truth)
\begin{scope}[shift={(8.7,-3.75)}]
 \fill[ctaskf, draw=ctask, line width=0.8pt, rounded corners=4pt] (-1.0,-1.15) rectangle (1.0,1.15);
 \begin{scope}[shift={(0,0.35)}]
  \draw[ctask, line width=0.8pt] (-0.65,-0.38) -- (0.65,-0.38);
  \fill[ctask!45] (-0.5,-0.38) rectangle (-0.22,0.02);
  \fill[ctask]    (-0.14,-0.38) rectangle (0.14,0.42);
  \fill[ctask!70] (0.22,-0.38) rectangle (0.5,0.2);
 \end{scope}
 \node[hd, ctask] at (0,-0.75) {Grader};
\end{scope}
\coordinate (Gw) at (7.7,-3.75); \coordinate (Ge) at (9.7,-3.75); \coordinate (Gs) at (8.7,-4.9);

% ================= REWARD (row 2, right; the output)
\begin{scope}[shift={(11.5,-3.75)}]
 \draw[creward, line width=1.2pt] (0,0.15) circle (0.66); \draw[creward, line width=0.5pt] (0,0.15) circle (0.53);
 \node[font=\small, creward] at (0,0.15) {$R_t$};
 \node[hd, creward] at (0,-0.85) {Score};
 \node[sub] at (0,-1.45) {edge recall: recovered\,/\,reference edges};
\end{scope}
\coordinate (Rw) at (10.84,-3.6);

% ================= GROUND TRUTH (below the grader, which alone reads it)
\begin{scope}[shift={(8.7,-6.15)}]
 \fill[cgtf, draw=cgt, line width=0.8pt, rounded corners=3pt] (-1.55,-0.55) rectangle (1.55,0.6);
 \gtgraph
 \node[font=\scriptsize\bfseries, cgt!70!black, align=center] at (0,-0.85) {Reference graph $G_t$};
 \node[font=\scriptsize, black!55] at (0,-1.13) {hidden from the solver};
\end{scope}
\coordinate (GTn) at (8.7,-5.55);

% ================= ARROWS
\draw[ar] (2.0,0) -- node[above, lab, align=center] {public\\ input $I_t$} node[pos=0.5, below=5pt, num] {1} (Aw);
\draw[ar] ([yshift=4.5mm]Ae) -- node[above, lab] {$a\in\mathsf{Act}$} node[pos=0.2, below=5pt, num] {2} ([yshift=4.5mm]Ew);
\draw[ar] ([yshift=-4.5mm]Ew) -- node[below, lab] {$o=\mathsf{Obs}_{\mathcal{E}}(a)$} node[pos=0.2, above=5pt, num] {3} ([yshift=-4.5mm]Ae);
\draw[ar] (As) -- (4.9,-2.1) -- node[pos=0.2, above=4pt, num] {4} (-1.2,-2.1) -- (Cn);
\draw[ar] (Ce) -- (Vw);
\draw[ar] (Ve) -- node[above, lab, fill=none] {$\mathbf{v}$} node[pos=0.5, below=5pt, num] {5} (Gw);
\draw[ar] (Ge) -- node[pos=0.5, below=5pt, num] {6} (10.84,-3.75);
\draw[dar] (GTn) -- (Gs);
\end{tikzpicture}%
}
\caption{Overview of an ABF-Bench run. (1)~The solver receives the
public input $I_t$ of task $t$. (2,\,3)~It queries the metered
environment and receives observations. (4)~It submits a report
$\mathcal{C}$ of operations, each citing the records that support it.
(5)~The verifier checks every operation against its cited records.
(6)~The grader scores the verdicts against the reference graph $G_t$.}
\label{fig:arch}
\end{figure}

Existing resources evaluate labeled transactions, protocol-specific
links, and incident reconstructions (Section~\ref{sec:related}). In
agentic blockchain forensics, LLM agents carry out the investigation
themselves~\cite{locard,risktagger}, and comparing their strategies
requires control over the records each solver sees and the cost of
acquiring them.

We introduce \textbf{ABF-Bench},\footnote{The benchmark will be released upon publication.} which evaluates evidence-backed
fund tracing through a shared, metered interface. Each task declares
seeds and a scope, and the solver submits operations together with the
records that support them. A verifier checks this support, and a
separate grader measures recovery of a hidden reference graph, so an
operation the records support remains valid even when the graph omits
it. We construct tasks
from incident addresses and ledger records and retain witnesses
that make each reference edge auditable. The released tasks combine multi-hop
transfers, DeFi exchanges, cross-chain round trips, custodial and
mixer deposits, and address poisoning.

Our contributions are:
\begin{enumerate}
\item We formulate evidence-backed blockchain fund tracing as an
interactive, verifiable agent task, opening blockchain forensics to
LLM agents.
\item We build ABF-Bench with a reusable case-to-task pipeline that
turns a real laundering incident into well-posed cross-chain tasks:
conservative rules determine each reference graph, and every reference
edge has a verified witness retrievable within the task's query
budget.
\item We propose the ABF agent harness, built around a working memory,
and evaluate it alongside general-purpose agents and fixed-order
tracers. Our harness outperforms both general-purpose agents in final
recall, including Codex on the same model, which shows that the task
rewards agent design. The harness also acquires evidence faster
than the tracers early in the budget and recognizes undisclosed
custody they walk past, at comparable validity. Yet every agent stops
early, so breadth-first tracing attains the highest final recall and
marks headroom for better agents.
\end{enumerate}

\section{Related Work}
\label{sec:related}
\label{sec:landscape}

\paragraph{Tracing and anti-money laundering.}
Bitcoin clustering and taint analysis infer relationships or allocate
commingled value under explicit assumptions~\cite{meiklejohn2013fistful,moser2014taint,tironsakkul2019taint}.
TRacer addresses account-based tracing~\cite{tracer}, and
Connector and ABCTracer associate activity across chains~\cite{connector,abctracer}.
AMLGuard studies incident-level flows and transaction
semantics~\cite{amlguard}. ABF-Bench evaluates both the
recovered movements and the acquisition of their supporting records,
starting from seeds under metered ledger access.

\paragraph{Datasets and agent evaluation.}
Elliptic and Elliptic2 provide labeled transactions and subgraphs
for detection~\cite{weber2019aml,elliptic2}. Protocol-level resources,
such as LOCARD's correspondence tasks and XSplicer's evidence-backed
crossings~\cite{locard,xsplicer}, evaluate links within specified
mechanisms. CryptoBench and AI-ABC cover broader
cryptocurrency-agent tasks~\cite{cryptobench,aiabc}.
Table~\ref{tab:positioning} compares these settings by scored object,
record access, and evidence checking.

\begin{table}[t]
\caption{Blockchain evaluation settings compared.}
\label{tab:positioning}
\centering\small
\begin{tabular}{@{}llll@{}}
\toprule
Setting & Scored object & Records & Evidence checked \\
\midrule
Labeled detection & labels & fixed dataset & no \\
Protocol linking & links in one mechanism & fixed dataset & per link \\
Incident flows & flows, accounts & fixed dataset & no \\
Crypto agent tasks & answers & tool calls & no \\
ABF-Bench & fund-flow operations & acquired under a budget & per operation \\
\bottomrule
\end{tabular}
\end{table}

\paragraph{Agentic blockchain forensics.}
Agentic blockchain forensics is a paradigm in which LLM agents carry
out forensic investigations on blockchains, deciding which records to
retrieve and how to interpret them. LOCARD formalizes the paradigm
and proposes a framework for it~\cite{locard}. RiskTagger applies
evidence-guided agents to post-incident money-laundering
analysis~\cite{risktagger}, and TxSum, TxRay, and FaultSeeker use
language models for transaction understanding and attack
analysis~\cite{txsum,txray,faultseeker}. These works build forensic
agents, and ABF-Bench supplies tasks on which such agents can be compared
with fixed-order tracers under a query budget. Our ABF agent
harness (Section~\ref{sec:harness}) belongs to this line.

\paragraph{Verifiable agent benchmarks.}
Agent benchmarks score runs against checks fixed before the run: test
suites in SWE-bench~\cite{swebench}, environment state in WebArena and
$\tau$-bench~\cite{webarena,taubench}, captured flags in
Cybench~\cite{cybench}, and short answers that are hard to find but
easy to verify in BrowseComp~\cite{browsecomp}. ABF-Bench brings this
design to forensics, where the check is the public ledger and the
answer is an evidence-backed fund-flow graph.

\section{Fund Tracing as a Verifiable Task}
\label{sec:claims}

This section defines the objects the benchmark builds on: the tracing
task (Definition~\ref{def:task}), the claims a solver reports
(Definition~\ref{def:report}), and the verdicts of the verifier
(Equation~\ref{eq:verdict}). Section~\ref{sec:tests} then locates the
difficulty of the task in four stages. Real-world attribution and
speculative deposit-to-withdrawal links lie outside the task.

\subsection{Tasks and Reports}
\label{sec:objects}

A declaration $S=(\Omega_S,\pi_S)$ consists of a scope and an
accounting convention. The scope $\Omega_S$ specifies the ledgers,
block windows, tracing direction, hop cap, traced assets, and stops.
The convention $\pi_S$ prescribes how commingled funds are
apportioned among earlier deposits, an apportionment that a recorded
transfer alone leaves open.

Let $\mathbb{L}$ denote the set of ledger records of the declared
ledgers within their block windows. Records are taken as final at the
declared confirmation depth, and each record is addressed by a
resolvable locator, so every investigator retrieves the same record
from the same locator. Membership in $\mathbb{L}$ fixes a record's
content, whereas its interpretation, including whether an apparent transfer
moves a traced asset, depends on the asset, the contract, and the
enclosing operation.

The set $O$ collects the admitted operation types, among them
transfers, exchanges, and cross-chain crossings, and assigns to each
a rule stating which records suffice to establish an operation of
that type. For example, an exchange is established by a receipt and
an execution trace, and a crossing by paired deposit and payout
records.

The evidence that establishes an operation is therefore determined by
the task, through the rules of $O$ applied to the records of
$\mathbb{L}$. Admitting other record classes, such as protocol
attestations or published internal records, would define other evidence
policies, whose trust may be incomparable. The current release admits
ledger records for all scored movements, and published incident
addresses enter only as assumptions on seeds and membership.

Seeds and claims are stated over \emph{fund states}. A fund state
$s=(\ell,a,x,p)$ comprises a ledger $\ell$, an account $a$, an asset
$x$, and a position $p$, the locator of the record at which the
holding is observed. Positions distinguish successive visits to the
same account, and $\mathcal{F}$ denotes the set of fund states.

\begin{definition}[Tracing task]
\label{def:task}
A tracing task is a tuple $t=(\mathbb{L},\Lambda,S,O)$ of admitted
records $\mathbb{L}$, a finite seed set $\Lambda\subseteq\mathcal{F}$,
a declaration $S$, and admitted operation types $O$.
\end{definition}

\begin{definition}[Claims and reports]
\label{def:report}
A claim $c=(s,s',\mathbf{q})$ asserts a movement of amounts
$\mathbf{q}$ from fund state $s$ to fund state $s'$, in their
respective assets. Its citation $A(c)$ is a finite set of locators of
records in $\mathbb{L}$. A report $\mathcal{C}=(C_1,\ldots,C_m)$
groups its claims into operations: each group $C_k$, $1\le k\le m$,
asserts one operation.
\end{definition}

\subsection{Verification}
\label{sec:verification}

Verification is separate from grading: the verifier judges each
operation against the records it cites, independently of the
reference graph, while the grader of Section~\ref{sec:metrics} scores
the resulting verdicts against that graph. For a group $C_k$, let
$A(C_k)=\bigcup_{c\in C_k}A(c)$ denote its citations and
$E_k=\mathbb{L}|_{A(C_k)}$ the records they locate. The deterministic
verifier returns
\begin{equation}
v_k=V(C_k,E_k;O)\in\{1,0,\bot\}.
\label{eq:verdict}
\end{equation}
The verdict is \emph{valid} ($1$) when an evidence rule in $O$ and
the records $E_k$ support every claim in $C_k$; \emph{invalid} ($0$)
when a claim contradicts $E_k$ or the group violates the required
output semantics; and \emph{indeterminate} ($\bot$) otherwise, in
particular when evidence is missing yet uncontradicted. A failed
retrieval therefore leaves an operation indeterminate and is silent on
whether the movement occurred. The solver's
operation-type label is advisory: verdicts rest on the claims and the
cited records alone.

\subsection{Why Tracing Is Hard}
\label{sec:tests}

Tracing is adversarial: launderers split flows across
accounts~\cite{meiklejohn2013fistful}, route them through exchanges,
mixers, and bridges~\cite{wang2023mixers,connector,abctracer}, and
plant look-alike addresses in the investigator's
view~\cite{tsuchiya2025poisoning}. Techniques change between
incidents: learned detectors degrade sharply after a single market
shutdown~\cite{weber2019aml}, so investigative effort concentrates
where a case departs from known patterns. Following the foraging and sensemaking loops observed in
intelligence analysis~\cite{pirolli2005sensemaking}, we describe an
investigation through four stages, with ledger queries as the
metered cost.
\emph{Search} chooses which records to acquire next and when to
stop: holdings fan out to many continuations, every query draws on
the same budget, and an uninspected branch silently truncates
everything downstream of it. \emph{Recognition} reads acquired
records as operations: a swap or a crossing appears only through
protocol-specific encodings, and an arrival at a custodial service
outside the declared stops must be recognized from behavior alone.
\emph{Evidence construction} assembles the records an operation's
rule requires, such as pairing a deposit with its payout across ledgers.
Partial evidence leaves a true movement indeterminate.
\emph{Reporting} decides which established operations enter the
submission: omissions forfeit recall, and unsupported additions are
invalid.

A fixed procedure settles each stage in advance, whereas an investigator
adapts to the case at hand. Section~\ref{sec:hyp} derives three
hypotheses from these stages.

\section{ABF-Bench}
\label{sec:bench}

ABF-Bench is the tuple
$\langle\mathcal{T},\mathcal{E},V,\sigma_{\mathrm{GT}}\rangle$: a
finite set $\mathcal{T}$ of tracing tasks (Definition~\ref{def:task}),
an evidence environment $\mathcal{E}$, the verifier $V$, and a
reference retrieval solver $\sigma_{\mathrm{GT}}$. Each task is
published with its instructions and budget, while its reference graph
and witnesses remain hidden. The reference solver builds a retrieval plan
from the reference graph, and the plan serves only to calibrate budgets.

\subsection{Design Goals}
\label{sec:goals}

Following the agentic benchmark checklist~\cite{zhu2025abc}, we design
for task validity, under which a task is solvable exactly when the
solver has the target capability, and outcome validity, under which a
score indicates task success. We add reproducibility.
\begin{description}[leftmargin=1.6em,itemsep=2pt,topsep=2pt]
\item[Task validity.] Every task is well-posed: its statement
determines the reference answer. Every reference edge passes the
verifier through its witness; every halt of the reference graph is
either stated in the task, as a declared stop or a boundary the
environment marks, or recognizable from the records, as undisclosed
custody is; and every witness can be retrieved within the budget. A
solver that follows the rules and recognizes every boundary therefore
attains full scores, as replaying the reference plan does. The halts
are conservative: tracing stops wherever the records leave a
continuation unresolved, so each task under-approximates the laundering
flow (Section~\ref{sec:gt}). Conversely, the reference graph stays
hidden, and a reference edge counts as recovered only when the solver
itself retrieved every record it cites (Section~\ref{sec:metrics}).
\item[Outcome validity.] A score reflects evidence-backed recovery.
The deterministic verifier judges each operation against the records
it cites, independently of the reference graph, and accepts any
evidence that satisfies a rule of $O$, so a valid alternative to the
reference witness passes (Section~\ref{sec:verification}).
\item[Reproducibility.] Each record is fixed by its locator, and the
verifier and grader are deterministic. Every solver receives the same
public input, query costs, and budget, and the budget is calibrated to
the reference plan alone (Sections~\ref{sec:task}
and~\ref{sec:metrics}).
\end{description}

\subsection{Interaction Under a Budget}
\label{sec:task}
\label{sec:environment}

The public input $I_t$ comprises the seeds, block windows, hop cap,
followed bridges, disclosed stops, available actions, submission
rules, and the budget. The reference graph, the reference
plan, and any marking of undisclosed custody stay hidden from
solvers.

The environment $\mathcal{E}=(\mathbb{L},\mathsf{Act},\kappa)$
provides ledger-reading queries $\mathsf{Act}$ with cost
$\kappa:\mathsf{Act}\to\mathbb{N}_0$. Queries return activity listings,
summaries, balances, transactions, receipts, traces, Bitcoin output
spends, contract-call results, and events. We write
$o=\mathsf{Obs}_{\mathcal{E}}(a)$ for the response to query $a$.
Observations retain native ledger structure and integer base units.
Records at which tracing halts, other than custody, carry a boundary
mark (Section~\ref{sec:gt}).
Each well-formed query costs $\kappa(a)=1$, and each page of an
activity listing counts as one query. Malformed calls, submission, and
the solver's internal computation are free, and submission ends the run.

As in prior work on agentic blockchain forensics~\cite{locard},
we represent a solver as a possibly randomized policy $\sigma$.
Given $I_t$ and the query history $h_n=((a_i,o_i))_{i=1}^n$, it
selects
\begin{equation}
u_{n+1}\sim\sigma(\cdot\mid I_t,h_n),
\label{eq:solver}
\end{equation}
either a query in
$\mathsf{Act}$ or $\operatorname{Submit}(\mathcal{C})$.
Internal computation and working memory belong to the policy, and a
deterministic tracer is the special case in which every choice is a point mass.
Every query prefix must satisfy
\begin{equation}
\sum_{i=1}^{n}\kappa(a_i)\leq B(t).
\label{eq:querybudget}
\end{equation}
Both agents and tracers know costs and budgets in advance, and every
admitted record can be retrieved.

Each report group $C_k$ describes one semantic operation, so a group of
arbitrary internal transfers is invalid. A router-mediated
swap, for example, must identify where the exchanged value comes to
rest, with evidence linking its input and output. Net balance changes
alone leave that link unproven. All claims of a group inherit its
verdict for scoring, so a genuine internal transfer submitted as a group
of its own can be invalid.

\subsection{The Reference Graph}
\label{sec:gt}
\label{sec:operation}

Tracing follows admitted assets and operations to declared boundaries:
it halts at custody and at unsupported bridge continuations,
and a mixer deposit establishes entry while leaving the exit open.
Some boundaries are disclosed. The others stay undisclosed and test
recognition, and we retain such a boundary only when its status can be
inferred from the records and the task rules settle its continuation.

Purity, the requirement that a holding contain traced value only,
delimits the traced region and keeps the reference graph conservative.
On account-based ledgers, a holding enters the region when traced value
arrives while it is empty and leaves it when outside value arrives. On
Bitcoin, the test applies to a transaction's non-dust inputs.
Commingled flows outside the region remain unresolved.

Let $\mathcal{K}_t$ denote the set of verifier-supported claims
reachable from $\Lambda$ within $S$. The finite reference graph
satisfies $G_t\subseteq\mathcal{K}_t$ and has a witness for each edge.
All compared methods use the same reference graphs and
verifier, and any amendment requires rescoring the affected methods.

\subsection{Scoring}
\label{sec:metrics}

The \emph{grader} in Figure~\ref{fig:arch} maps the report
$\mathcal{C}$, operation verdicts $\mathbf{v}=(v_1,\ldots,v_m)$,
and reference graph $G_t$ to edge recall $R_t$, fidelity $F_t$, and
edge precision $P_t$, passing verdicts through unchanged.
An edge is identified by its anchoring transaction and its target fund
state in canonical form, so two descriptions of one landing denote one
edge, and $|G_t|$ counts distinct edges. A reference edge is recovered
when the report contains it in a valid operation and the run retrieved
every record it cites. Let
$M_t(\mathcal{C},\mathbf{v})\subseteq G_t$ denote the set of
recovered reference edges. This
retrieval condition enforces the instruction to cite only served
records. The verifier resolves every citation, including records the
solver left unretrieved, so a lenient count without the condition can
exceed $R_t$ (Section~\ref{sec:steps}). Formally,
\begin{align}
\operatorname{Grade}_t(\mathcal{C},\mathbf{v};G_t)&=(R_t,F_t,P_t),
\label{eq:grader}\\
R_t&=\frac{|M_t(\mathcal{C},\mathbf{v})|}{|G_t|},
\qquad |G_t|>0.
\label{eq:recall}
\end{align}
Edge recall $R_t$ is the primary score, and valid downstream edges count
as recovered even when upstream operations are rejected. We report
macro-averaged recall, with failed solver runs kept in the average.

Fidelity $F_t$ is the fraction of submitted operations passing verification,
so it judges the cited records against the ledger and leaves the
retrieval condition to $R_t$.
Edge precision $P_t$ divides the distinct reference edges of valid
operations by the number of submitted edges, including out-of-scope
edges, so a valid edge that continues past a halt of the task lowers
it. Like fidelity, it is computed from the verdicts alone and leaves
the retrieval condition to $R_t$. Either metric is
undefined when its denominator is zero.

Let $\eta(t)$ count the reference solver's actions to retrieve the
reference graph's records, and $\delta(t)$ count one activity
listing per distinct source holding. The query budget, set from the
reference plan alone, is
\begin{equation}
B(t)=\left\lceil 2\bigl(\eta(t)+\delta(t)\bigr)\right\rceil.
\label{eq:budget}
\end{equation}
Other solvers may reuse evidence differently or need additional
listing pages. Budgets stay
fixed after reference-graph amendments, and on every retained task
the reference plan still fits within $B(t)$
(Section~\ref{sec:setup}). Query count measures evidence access, and
model tokens and monetary costs are reported separately.

\section{From Incident to Benchmark}
\label{sec:gen}
\label{sec:generator}

Given ledger records $\mathbb{L}$, published incident addresses
$\Psi$, a declaration $S$, and operation rules $O$, we construct
\begin{equation}
\operatorname{Gen}(\mathbb{L},\Psi,S,O)
=\langle\mathcal{T},\mathcal{E},V,\sigma_{\mathrm{GT}}\rangle.
\label{eq:gen}
\end{equation}
Published addresses, with reviewed amendments, determine the seed fund
states $\Lambda$.

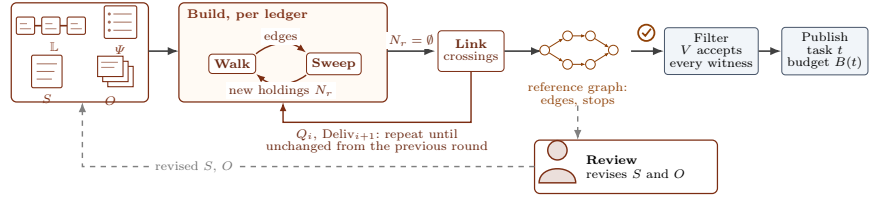
\begin{figure}[t]
\centering
\resizebox{0.93\linewidth}{!}{%
\begin{tikzpicture}[font=\small,
  gbox/.style={draw=cgen, fill=cgenf!60, rounded corners=3pt, line width=0.7pt},
  gstep/.style={draw=cgen, fill=white, rounded corners=2pt, font=\scriptsize\bfseries, inner sep=3pt, align=center, text=cgen!85!black},
  tstep/.style={draw=ctask, fill=ctaskf, rounded corners=2pt, font=\scriptsize, inner sep=3pt, minimum width=1.2cm, align=center},
  loop/.style={-{Latex[length=1.8mm]}, line width=0.8pt, cgen},
  guard/.style={font=\scriptsize, cgen!85!black, align=center, inner sep=1.5pt, fill=white}]
% ---- inputs (one card)
\fill[white, draw=cgen, rounded corners=3pt, line width=0.7pt] (0,-0.95) rectangle (2.7,1.0);
\begin{scope}[shift={(0.7,0.45)}] \chain{-0.6}{0.1}{3}{cgen} \node[sub, cgen, anchor=north] at (0.15,-0.12) {$\mathbb{L}$}; \end{scope}
\begin{scope}[shift={(2.15,0.62)}] \listicon \node[sub, cgen, anchor=north] at (0,-0.34) {$\Psi$}; \end{scope}
\begin{scope}[shift={(0.7,-0.35)}] \scrollicon \node[sub, cgen, anchor=north] at (0,-0.34) {$S$}; \end{scope}
\begin{scope}[shift={(2.15,-0.45)}] \templates \node[sub, cgen, anchor=north] at (-0.2,-0.28) {$O$}; \end{scope}
\draw[ar] (2.7,0.05) -- (3.3,0.05);
% ---- per-ledger build: Walk <-> Sweep loop
\fill[gbox] (3.3,-0.95) rectangle (7.4,1.0);
\node[cgen, anchor=north west, font=\scriptsize\bfseries] at (3.35,0.98) {Build, per ledger};
\node[gstep] (walk) at (4.35,-0.2) {Walk};
\node[gstep] (sweep) at (6.35,-0.2) {Sweep};
\draw[loop] (walk.north east) to[bend left=38] node[guard, above, yshift=-1pt] {edges} (sweep.north west);
\draw[loop] (sweep.south west) to[bend left=38] node[guard, below, yshift=1pt] {new holdings $N_r$} (walk.south east);
\draw[ar] (7.4,0.05) -- node[guard, above, yshift=1pt] {$N_r=\emptyset$} (8.35,0.05);
% ---- cross-ledger link, closing the outer loop
\node[gstep, minimum height=8mm, minimum width=1.3cm] (link) at (9.05,0.05) {Link\\[-1pt]\scriptsize\mdseries crossings};
\draw[loop] (link.south) -- (9.05,-1.35) -- (5.35,-1.35) -- (5.35,-0.95);
\node[guard, anchor=north] at (7.2,-1.42) {$Q_i$, $\mathrm{Deliv}_{i+1}$: repeat until\\[-1pt]unchanged from the previous round};
\draw[ar] (link.east) -- (10.45,0.05);
% ---- reference graph
\begin{scope}[shift={(11.3,0.05)}] \gtgraph \end{scope}
\node[sub, cgt!80!black, anchor=north, align=center] at (11.15,-0.45) {reference graph:\\[-1pt]edges, stops};
% ---- publish
\draw[ar] (12.2,0.05) -- (12.85,0.05);
\node[tstep, anchor=west] (filter) at (12.85,0.05) {Filter\\[-1pt]$V$ accepts\\[-1pt]every witness};
\draw[ar] (filter.east) -- ++(0.45,0) coordinate (pubw);
\node[tstep, anchor=west] (publish) at (pubw) {Publish\\[-1pt]task $t$\\[-1pt]budget $B(t)$};
% ---- review card (people), closes the loop on S and O
\fill[white, draw=cgen, rounded corners=3pt, line width=0.7pt] (10.3,-2.75) rectangle (13.9,-1.7);
\begin{scope}[shift={(10.75,-2.22)}] \person{cgen} \end{scope}
\node[font=\scriptsize, align=left, anchor=west] at (11.2,-2.22) {\textbf{Review}\\ revises $S$ and $O$};
\draw[dar] (11.15,-1.02) -- (11.15,-1.7);
\draw[dar] (10.3,-2.22) -- (1.4,-2.22) -- (1.4,-0.95);
\node[guard, text=black!55] at (3.6,-2.22) {revised $S$, $O$};
\end{tikzpicture}%
}
\caption{Building ABF-Bench from an incident. An inner loop per ledger
and an outer loop across ledgers build the incident's reference graph,
and human review revises $S$ and $O$ between builds. Each task is a
seed-rooted region of this graph and is published only if $V$ accepts
the witness of every edge it retains.}
\label{fig:gen}
\end{figure}

\begin{table}[t]
\caption{Composition of the evaluated tasks; categories
overlap.}
\label{tab:suite}
\centering\small
\setlength{\tabcolsep}{3pt}
\begin{tabular*}{\linewidth}{@{\extracolsep{\fill}}lrrr@{}}
\toprule
\multicolumn{2}{@{}l}{\emph{Scale and budget}} & Total & Per-task range \\
\midrule
\multicolumn{2}{@{}l}{Evaluated tasks} & 31 & -- \\
\multicolumn{2}{@{}l}{Reference edges} & 2044 & 20--107 \\
\multicolumn{2}{@{}l}{Largest holding fan-out} & -- & 4--28 \\
\multicolumn{2}{@{}l}{Query budget $B(t)$} & -- & 98--510 \\
\midrule
\emph{Operations and ledger coverage} & \shortstack[r]{Tasks with\\feature} & \shortstack[r]{Total\\edges} & \shortstack[r]{Edges\\per task} \\
\addlinespace[2pt]
Transfers & 31 & 1658 & 13--99 \\
Exchanges through venues & 28 & 166 & 0--21 \\
Cross-chain crossings & 31 & 220 & 1--24 \\
Bitcoin edges (UTXO) & 31 & 503 & 1--59 \\
\midrule
\multicolumn{4}{@{}l}{\emph{Custody boundaries}} \\
Into undisclosed custody & 23 & 113 & 0--13 \\
Into disclosed custody & 9 & 41 & 0--13 \\
\bottomrule
\end{tabular*}

\end{table}

Two nested iterations resolve dependencies within and across ledgers
(Figure~\ref{fig:gen}).

\paragraph{Within-ledger iteration.}
On Ethereum, $\operatorname{Walk}$ expands the current frontier, and
$\operatorname{Sweep}$ replays ledger records to determine which
holdings contain traced value and during which intervals. Replay can expose holdings
outside the discovered paths or extend their tracing windows, and the
set $N_r$ of these expansions after round $r$ feeds the next walk. The inner
loop stops when $N_r=\emptyset$ and all membership decisions are
resolved. Bitcoin replaces the walk--sweep loop with transaction
replay under its input-membership rules.

\paragraph{Across-ledger iteration.}
$\operatorname{Link}$ binds evidenced deposits and payouts and supplies
entry records on the destination ledger. In outer round $i$, linking
the Ethereum build's movements yields Bitcoin entries $Q_i$, and Bitcoin
replay and linking yield return deliveries $\mathrm{Deliv}_{i+1}$.
These deliveries can retroactively change Ethereum membership and
force a fresh build, which may expose further crossings. Starting from
$\mathrm{Deliv}_0=\emptyset$, the stopping check requires successive
rounds to coincide:
$(Q_i,\mathrm{Deliv}_{i+1})=(Q_{i-1},\mathrm{Deliv}_i)$ for
$i\geq1$.
A round that repeats an older pair signals oscillation instead.

\paragraph{Human review and task publication.}
Outside the two iterations, human reviewers audit the resulting graph
and its evidence, and record seed, boundary, and rule revisions for
subsequent builds. Two independent readings of the rules test them in
practice: the fixed-order tracers, implemented from the public task
specification separately from the generator and verifier, and the
evaluated agents. Every one of these solvers has at least
\resFidMinPct\% of its operations accepted by the verifier (Table~\ref{tab:results}).

We select seed-rooted regions for operation diversity, crossings,
boundaries, and branching, without regard to solver performance, and cap
pairwise overlap at 25\% of the edges of the smaller reference graph. The release filter excludes tasks that depend on
unsettled interpretation rules and requires $V$ to accept a witness
for every retained reference edge. Each task receives the shared evidence interface and a
query budget derived from the reference plan (Section~\ref{sec:metrics}).

\subsection{The Released Task Suite}
\label{sec:case}

The case is the Bybit theft of February 2025, followed for 90 days
across Ethereum and Bitcoin. Seeds derive from the FBI notice and
Bybit's published incident addresses~\cite{fbi2025psa,hackscan2025}.
The retained region combines native and token transfers, DeFi
exchanges and routers, and cross-chain round trips. Mixer and
custodial-service deposits are endpoints, except where the admitted
records establish a continuation. Poisoning and dust activity remain visible
in the observations.

Of the 58 tasks produced by the authoring pass, 33 pass the release
filter and are published. The construction interface separates
case-specific seeds and boundaries from the tracing procedure. A new
case on supported ledgers supplies its own seeds, boundaries, and
declaration, while a new ledger or operation type also requires evidence
rules and implementation support.

The released tasks exercise all four stages of
Section~\ref{sec:tests} (Table~\ref{tab:suite}). One holding fans out
to as many as \suiteMaxOutDegMax{} reference edges, so following a
single continuation leaves work unfinished. Tasks disclose
\suiteRosterDisclosed{} custodians and leave the rest to be recognized.
The reference graphs stop at \suiteHiddenArrivals{} edges
into undisclosed custody, nearly three times the number into
disclosed custody.

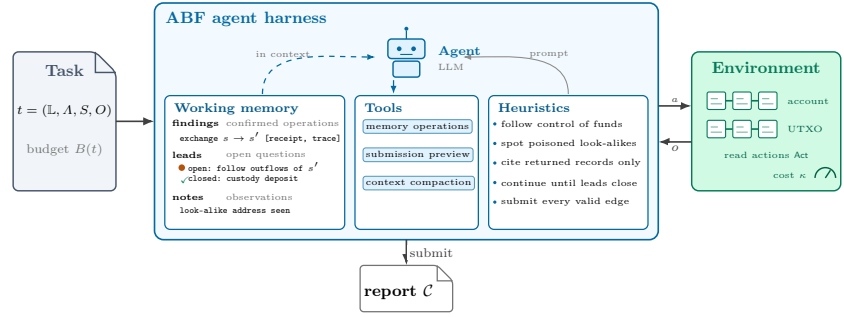
\begin{figure}[t]
\centering
\resizebox{0.90\linewidth}{!}{%
\begin{tikzpicture}[font=\small,
  mod/.style={draw=csolver, fill=white, rounded corners=3pt, line width=0.7pt},
  mh/.style={font=\scriptsize\bfseries, text=csolver!85!black, anchor=north west},
  li/.style={font=\scriptsize, anchor=west, inner sep=1pt},
  chip/.style={draw=csolver!70, fill=csolverf, rounded corners=1.5pt, font=\tiny\ttfamily, inner sep=1.6pt},
  tagd/.style={font=\tiny, text=black!55}]

% ================= TASK
\fill[ctaskf, draw=ctask, line width=0.8pt, rounded corners=3pt] (-3.2,-1.3) -- (-1.2,-1.3) -- (-1.2,1.0) -- (-1.6,1.4) -- (-3.2,1.4) -- cycle;
\draw[ctask, line width=0.8pt] (-1.6,1.4) -- (-1.6,1.0) -- (-1.2,1.0);
\node[hd, ctask] at (-2.2,1.05) {Task};
\node[font=\scriptsize] at (-2.2,0.25) {$t=(\mathbb{L},\Lambda,S,O)$};
\node[font=\scriptsize, text=black!55] at (-2.2,-0.5) {budget $B(t)$};
\draw[ar] (-1.2,0.05) -- (-0.45,0.05);

% ================= HARNESS container
\fill[csolverf!40, draw=csolver, line width=0.9pt, rounded corners=5pt] (-0.45,-2.25) rectangle (9.35,2.35);
\node[hd, csolver, anchor=north west] at (-0.35,2.3) {ABF agent harness};

% agent (robot) in the centre top
\begin{scope}[shift={(4.45,1.35)}]
 \draw[csolver, line width=0.7pt] (0,0.3) -- (0,0.46); \fill[csolver] (0,0.5) circle (1.1pt);
 \fill[csolverf, draw=csolver, line width=0.8pt, rounded corners=2pt] (-0.36,-0.08) rectangle (0.36,0.3);
 \fill[csolver] (-0.14,0.11) circle (1.3pt); \fill[csolver] (0.14,0.11) circle (1.3pt);
 \draw[csolver, line width=0.6pt] (-0.12,0.0) -- (0.12,0.0);
 \fill[csolverf, draw=csolver, line width=0.8pt, rounded corners=1.5pt] (-0.26,-0.44) rectangle (0.26,-0.13);
 \node[font=\scriptsize\bfseries, text=csolver, anchor=west] at (0.5,0.05) {Agent};
 \node[font=\tiny, text=black!55, anchor=west] at (0.5,-0.25) {LLM};
\end{scope}

% ---- module 1: Working memory
\fill[mod] (-0.25,-2.05) rectangle (3.25,0.55);
\node[mh] at (-0.18,0.52) {Working memory};
\node[li, font=\tiny\bfseries] at (-0.15,0.0) {findings};
\node[li, font=\tiny, text=black!55] at (0.9,0.0) {confirmed operations};
\node[li, font=\tiny\ttfamily] at (0.0,-0.26) {exchange $s\to s'$ [receipt, trace]};
\node[li, font=\tiny\bfseries] at (-0.15,-0.6) {leads};
\node[li, font=\tiny, text=black!55] at (0.9,-0.6) {open questions};
\fill[cgt] (0.08,-0.85) circle (1.8pt); \node[li, font=\tiny\ttfamily] at (0.16,-0.85) {open: follow outflows of $s'$};
\node[li, font=\tiny, text=cenv] at (0.0,-1.08) {\checkmark}; \node[li, font=\tiny\ttfamily] at (0.16,-1.08) {closed: custody deposit};
\node[li, font=\tiny\bfseries] at (-0.15,-1.42) {notes};
\node[li, font=\tiny, text=black!55] at (0.9,-1.42) {observations};
\node[li, font=\tiny\ttfamily] at (0.0,-1.68) {look-alike address seen};

% ---- module 2: Tools
\fill[mod] (3.45,-2.05) rectangle (5.85,0.55);
\node[mh] at (3.52,0.52) {Tools};
\node[chip, anchor=west, font=\tiny] at (3.6,-0.05) {memory operations};
\node[chip, anchor=west, font=\tiny] at (3.6,-0.6) {submission preview};
\node[chip, anchor=west, font=\tiny] at (3.6,-1.15) {context compaction};

% ---- module 3: Heuristics
\fill[mod] (6.05,-2.05) rectangle (9.15,0.55);
\node[mh] at (6.12,0.52) {Heuristics};
\foreach \y/\t in {0.0/follow control of funds,-0.38/spot poisoned look-alikes,-0.76/cite returned records only,-1.14/continue until leads close,-1.52/submit every valid edge} {
  \node[li, font=\tiny] at (6.25,\y) {\t}; \fill[csolver] (6.17,\y) circle (0.9pt);}

% agent <-> modules
\draw[-{Latex[length=1.6mm]}, csolver, line width=0.7pt] (4.2,0.85) -- (4.2,0.58);
\draw[-{Latex[length=1.6mm]}, csolver, line width=0.7pt, dashed] (1.65,0.58) to[out=90, in=180] node[tagd, above, pos=0.35] {in context} (3.85,1.3);
\draw[-{Latex[length=1.6mm]}, black!45, line width=0.6pt] (7.6,0.58) to[out=90, in=0] node[tagd, above, pos=0.35] {prompt} (5.55,1.3);

% ================= ENVIRONMENT
\fill[cenvf, draw=cenv, line width=0.8pt, rounded corners=4pt] (10.0,-1.3) rectangle (12.9,1.4);
\node[hd, cenv] at (11.45,1.1) {Environment};
\chain{10.3}{0.45}{3}{cenv} \node[font=\tiny, cenv!80!black, anchor=west] at (11.75,0.45) {account};
\chain{10.3}{-0.1}{3}{cenv} \node[font=\tiny, cenv!80!black, anchor=west] at (11.75,-0.1) {UTXO};
\node[font=\tiny, cenv!80!black] at (11.45,-0.6) {read actions $\mathsf{Act}$};
\draw[cenv!80!black, line width=0.8pt] (12.4,-1.05) arc (180:0:0.2); \draw[cenv!80!black, line width=0.8pt] (12.6,-1.05) -- (12.7,-0.9);
\node[font=\tiny, cenv!80!black, anchor=east] at (12.35,-1.0) {cost $\kappa$};
\draw[ar] (9.35,0.35) -- node[lab, above, font=\tiny] {$a$} (10.0,0.35);
\draw[ar] (10.0,-0.35) -- node[lab, below] {$o$} (9.35,-0.35);

% ================= OUTPUT: the report
\draw[ar] (4.45,-2.25) -- node[lab, fill=white, right] {submit} (4.45,-2.75);
\fill[white, draw=black!55, line width=0.7pt, rounded corners=2pt] (3.55,-3.65) -- (5.35,-3.65) -- (5.35,-3.05) -- (5.05,-2.75) -- (3.55,-2.75) -- cycle;
\draw[black!55, line width=0.7pt] (5.05,-2.75) -- (5.05,-3.05) -- (5.35,-3.05);
\node[font=\small\bfseries] at (4.3,-3.27) {report $\mathcal{C}$};
\end{tikzpicture}%
}
\caption{The ABF agent harness equips an LLM agent with three
modules: working memory, tools, and heuristics. The agent queries the
same environment as every other solver and submits a report.}
\label{fig:abf}
\end{figure}

\section{The ABF Agent Harness}
\label{sec:harness}

We propose the ABF agent harness, which equips an LLM agent
for forensic investigation with three modules (Figure~\ref{fig:abf}).
The agent makes every tracing decision, choosing each query and
interpreting each record.
\begin{description}[leftmargin=1.2em,itemsep=0pt,topsep=2pt]
\item[Working memory.] A working memory~\cite{hu2025memory},
maintained by the agent itself~\cite{zhang2026landscape}, that represents
the progress of the investigation. It comprises three sections:
\emph{findings}, the confirmed operations that constitute the report;
\emph{leads}, the open questions still to pursue, each closed with its
disposition; and \emph{notes}, observations kept for later use.
\item[Tools.] Operations on the working memory, a submission preview
that checks the report before it is submitted, and context compaction
by summarization.
\item[Heuristics.] Five short rules in the prompt. Two encode
blockchain-forensics knowledge: follow the control of funds over the
movement of coins, and check assets and full addresses against
poisoned look-alikes. One grounds the report: every cited value comes
from a returned record or the task text. Two govern stopping and
reporting: continue until every open lead is exhausted or out of
scope, or the budget is spent, and submit every edge found valid.
\end{description}
The ABF agent harness targets blockchain forensics in general,
independent of any case, protocol, or service, and the varied tasks of
Section~\ref{sec:case} test this generality. The design is inspired by the LOCARD
framework~\cite{locard}.

\section{Experiments}
\label{sec:experiments}

The experiments test three hypotheses about agents in fund tracing
(Section~\ref{sec:hyp}). They compare solvers on edge recall, the
primary score, and on fidelity and edge precision
(Section~\ref{sec:metrics}), and report the query and token cost of
each run.

\subsection{Hypotheses}
\label{sec:hyp}

We expect the generality of language models to help forensic agents
where an investigation depends on the specific case, at little cost to
the reliability of their evidence and reports: such models perform new
tasks specified only by instructions or a few
examples~\cite{brown2020gpt3}. Over the stages of
Section~\ref{sec:tests}, we state three hypotheses.
\begin{description}[leftmargin=1.6em,itemsep=0pt,topsep=2pt]
\item[H1 (efficiency).] In the search stage, agents may uncover
evidence faster than rule-based methods.
\item[H2 (adaptability).] In the recognition and evidence construction
stages, agents may identify operations and custody and construct their
evidence from the records alone.
\item[H3 (reliability).] In the evidence construction and reporting
stages, agents may be nearly as reliable as rule-based methods.
\end{description}
Section~\ref{sec:steps} measures efficiency by evidence recall
against queries spent, adaptability by the operations and
undisclosed custody that agents identify, and reliability by fidelity
and by the share of retrieved transfers left unreported. The
fixed-order tracers of Section~\ref{sec:solvers} instantiate the
rule-based methods.

\subsection{Solvers}
\label{sec:solvers}

Three solver families share one evidence interface and receive the
same public input $I_t$ and query costs, and the reference graph stays
hidden from all of them.

\paragraph{Fixed-order tracers.}
The fixed-order tracers are implemented against the public
interface, independently of the generator and verifier code. Each
expands outgoing activity under the declared convention $\pi_S$,
handles recognized operations and crossings, and follows a fixed
frontier order until it finishes or exhausts its budget. We run three
orders: \texttt{bfs} (breadth-first), \texttt{dfs} (depth-first), and
\texttt{greedy} (largest incoming amount first). The best-in-hindsight
tracer, the best of the three on each task at each budget point, serves
only as an analysis reference.

\paragraph{General-purpose agents.}
We evaluate the Codex and Claude Code command-line harnesses,
restricted to the benchmark tools. Their pretrained models retain
general knowledge, and they receive only the public input $I_t$.

\paragraph{ABF agent harness.}
The harness of Section~\ref{sec:harness} runs the same model and
reasoning effort as Codex, so the comparison with Codex holds the
model fixed and varies the harness.

\subsection{Setup}
\label{sec:setup}

The comparison uses \resSuite{} of the 33 released tasks: one task
was excluded after repeated environment failures for one arm, and a
second after its amended reference graph outgrew its published
reference-plan budget.
All methods are scored against the released reference graphs under the
budgets published with the tasks.

Each isolated container accesses only the evidence interface and its
model endpoint, and all queries, observations, and submissions are
logged.
The fixed-order tracers run under a 2\,GB memory limit. Codex and the
ABF agent harness use \texttt{gpt-5.5} at high reasoning effort, and
Claude Code uses \texttt{claude-opus-4-8} at high effort. Headline
results use the first round, and a second round covers \resAbfPairN{}
harness tasks and \resCodexPairN{} Codex tasks.

Solver failures remain in the suite with zero recall for missing
submissions, and we also report finished-run means. Tracer failures come
from memory-limit exhaustion and timeouts. Trials ended by
environment failures were rerun (four harness, two Codex, and four
Claude Code trials on the evaluated tasks), and reruns after a failure
of the environment itself used its repaired version.

\subsection{Main Results}
\label{sec:results}

\begin{table}[t]
\caption{Main results, first round, means per task. Budget used is the
fraction of $B(t)$ spent; cost is model API cost at list price. Every
agent's recall comes at under half the budget spent; the tracers spend
nearly all of it.}
\label{tab:results}
\centering\small
\resizebox{\linewidth}{!}{\begin{tabular}{@{}lrrrrrr@{}}
\toprule
Solver & \textbf{Recall} & Fidelity & Edge precision & Budget used & Output tokens & Cost \\
\midrule
\multicolumn{7}{@{}l}{\emph{Agents}} \\
\rowcolor{csolverf}ABF agent harness (ours; GPT-5.5) & \textbf{0.75} & \textbf{0.98} & \textbf{0.92} & 0.47 & 43k & \$9.4 \\
Codex (GPT-5.5) & 0.47 & \textbf{0.98} & \textbf{0.92} & 0.33 & 24k & \$2.7 \\
Claude Code (Opus 4.8) & 0.10 & 0.95 & 0.87 & 0.16 & 122k & \$6.5 \\
\midrule
\multicolumn{7}{@{}l}{\emph{Fixed-order tracers}} \\
Tracer (bfs) & \textbf{0.88} & 0.98 & \textbf{0.78} & 0.92 & -- & -- \\
Tracer (dfs) & 0.77 & \textbf{0.99} & \textbf{0.78} & 0.87 & -- & -- \\
Tracer (greedy) & 0.54 & 0.95 & 0.63 & 0.88 & -- & -- \\
\bottomrule
\end{tabular}
}
\end{table}

On edge recall, the primary score, the ABF agent harness outperforms
both general-purpose agents, and breadth-first tracing ranks highest
(Table~\ref{tab:results}). Claude Code's \resClaudeRecall{} is set by the
retrieval condition: scored on the verdicts alone, it reaches
\resClaudeRecallLenient{}, and Section~\ref{sec:steps} traces the gap to
citations of records it left unretrieved. The harness beats \texttt{bfs} on only
\resAbfBeatsBfs{} of \resSuite{} tasks. Tracer recall counts runs that
ended without a submission as zero, and excluding them raises
\texttt{bfs} only to \resBfsRecallFin{}, so the ranking holds. The
agents' edge precision is higher because the tracers follow service sweeps
past undisclosed custody into valid edges outside the reference graph
(Section~\ref{sec:steps}).

The benchmark rewards agent design and leaves headroom for it. The
harness and Codex share a model, so the gap between them
(\resAbfRecall{} against \resCodexRecall{}) comes from the design
around the model, and breadth-first tracing shows that at least
\resBfsRecall{} is attainable within the same budgets, while the harness
reaches its recall with \resAbfShare\% of its budget spent.

Every agent's recall falls with task size: from the smallest to the largest third
of tasks by reference edges, the harness drops from \resAbfSmall{} to
\resAbfLarge{}, and both general-purpose agents lose a larger share of
their recall.
Across tasks, the harness's recall correlates more strongly with the
largest fan-out of a single holding than with the number of reference
edges (Spearman $\rho=\resAbfRhoOutdeg$ against $\resAbfRhoEdges$), so
its losses are associated more with branching than with size.

Per-task results are noisier than the means: across repeated runs, the
within-task standard deviation of recall is \resAbfPairSD{} for the
harness and \resCodexPairSD{} for Codex, while mean recall agrees within
0.01 (Section~\ref{sec:setup}).

\subsection{Where Agents Gain and Lose}
\label{sec:steps}

We examine each hypothesis on the recorded query and submission
trajectories. The results characterize the evaluated solvers on this
suite. Whether the hypotheses hold for agents in general remains open.

\paragraph{H1: Efficiency.}
The harness acquires reference evidence faster than the fixed-order
tracers early in the budget, while the general-purpose agents trail them. We
measure acquisition by evidence recall. For a reference
claim $c$, let $D(c)$ denote its reference witness, the set of records
the reference graph retains for $c$. The witness may differ from the
submitted citations $A(c)$, and other records may also suffice. For a run on task
$t$, let $j_c$ be the cumulative query cost at which every record of
$D(c)$ has first been retrieved as a full record, beyond its listing
entry, or $\infty$ otherwise. Evidence recall at
query cost $j$ is
\begin{equation}
R^{\mathrm{ev}}_t(j)=\frac{|\{c\in G_t:j_c\leq j\}|}{|G_t|}.
\label{eq:anytime}
\end{equation}
Evidence recall counts a reference claim once its witness has been read, submitted
or unsubmitted. It therefore measures access to the reference records,
and its final value can differ from edge recall when reports omit
retrieved evidence or rely on other sufficient witnesses.

\begin{figure}[t]
\centering
\includegraphics[width=0.88\linewidth]{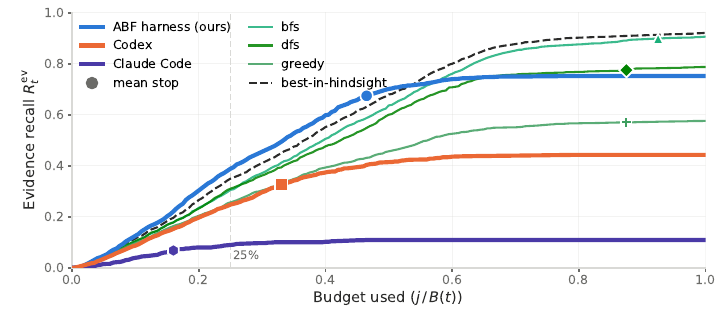}
\caption{Evidence recall $R^{\mathrm{ev}}_t$ against budget used
$j/B(t)$, averaged over tasks. Markers show each solver's mean stopping
point.}
\label{fig:anytime}
\end{figure}

\begin{figure}[t]
\centering
\includegraphics[width=0.88\linewidth]{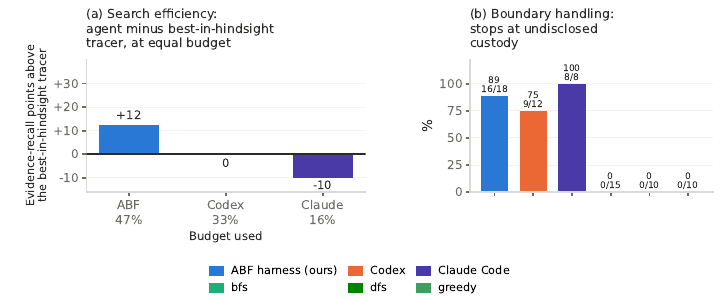}
\caption{Search efficiency and boundary handling. (a)~H1: evidence
recall at each agent's own stopping point minus that of the
best-in-hindsight tracer at the same budget, averaged over tasks. (b)~H2: stops at
reached undisclosed custody points whose outflows the run listed and
from which it could still continue.}
\label{fig:steps}
\end{figure}

The harness outpaces even the best-in-hindsight tracer
(Figure~\ref{fig:anytime}). At a
quarter of the budget, the harness reaches an evidence recall of
\resAbfPlainQuarter\% against \resEnvQuarter\% for this reference and
still leads at half the budget, while both general-purpose agents trail
(Table~\ref{tab:steps}). The harness also leads at its own stopping
point on \resAbfAheadOwnStop{} of \resSuite{} tasks, while Codex runs
level with the best-in-hindsight tracer and Claude Code falls behind
(Figure~\ref{fig:steps}(a)).

The early lead fades because every agent stops early: the
best-in-hindsight tracer overtakes the harness after
\resAbfEnvCross\% of the budget, and even the harness leaves
\resAbfUnused\% of its budget unused on average
(Table~\ref{tab:results}). Transcripts show Codex declining further
outflows as too costly with two thirds of its budget left, harness
runs closing their open leads and submitting, and Claude Code
declaring completion after 17 of 428 actions. The missing evidence
lies in branches the agents left unexamined, which hold a quarter of
all reference edges for the harness and more than half for Codex. What
continuing would recover remains unmeasured. H1 holds for the
harness early in the budget, and its final evidence recall falls
behind because it stops early.

\paragraph{H2: Adaptability.}
The harness recognizes operations about as often as the tracers, and
all agents identify undisclosed custody far more often. Given the
diagnostic records, the harness claims \resAbfOpsClaimed\% of retrieved
operations against \resBfsOpsClaimed\% for \texttt{bfs}
(Table~\ref{tab:steps}), rates that combine interpretation, evidence
construction, and reporting. Codex claims only
\resCodexClaimedCrossing\% of retrieved crossings, failing to carry the
deposit memo to the Bitcoin payout. \texttt{bfs} reads bridge-deposit
records but lacks a rule to emit them, whereas the harness submits all
of them, which shows recognition beyond the rules of \texttt{bfs} within the
operation types of this suite.

Custody is where the methods differ. \texttt{bfs} stops at disclosed
custody and at boundaries the environment marks, but at an undisclosed
deposit it may follow the service's sweep (Figure~\ref{fig:steps}(b)).
We count only points whose outflows the run itself listed and from
which it could still continue, with budget left and below the hop cap.
Listing separates inspection of the frontier from mere arrival. Each
agent stops at three quarters or more of these points and the
fixed-order tracers at none: every tracer stop came with its budget
spent. Custody recognition thus supports H2. Following sweeps, \texttt{bfs} emits
\resBfsTpPast{} valid edges beyond undisclosed custody, most of its
unmatched output, yet it reaches more reference edges into these points
(\resBfsTpArr\% against \resAbfTpArr\% for the harness): reaching
custody and recognizing it favor different methods.

The agents' custody decisions may also draw on case knowledge from
pre- or post-training. The served text leaves out both names of the
case: Bybit, the victim exchange, and Lazarus, the North Korean group
to which the theft is attributed. Yet Claude Code names the incident
in 21 of 31 surveyed transcripts, against 2 of 33 for Codex and 0 of 33 for the
harness. The survey covers runs beyond the common suite. A mention
establishes only that the knowledge is available, and silence leaves its
presence open.

\begin{table}[t]
\caption{Diagnostics by hypothesis, first round. Row denominators differ.}
\label{tab:steps}
\centering\small
\setlength{\tabcolsep}{3pt}
% Presentation view of the generated results/numbers.tex macros.
\begin{tabular}{@{}lrrrrrr@{}}
\toprule
Measure (\%) & ABF & Codex & Claude & \texttt{bfs} & \texttt{dfs} & \texttt{greedy} \\
\midrule
\multicolumn{7}{@{}l}{\emph{H1: evidence recall}} \\
At 25\% of budget & \resAbfPlainQuarterCell & \resCodexPlainQuarterCell & \resClaudePlainQuarterCell & \resBfsPlainQuarterCell & \resDfsPlainQuarterCell & \resGreedyPlainQuarterCell \\
At 50\% of budget & \resAbfPlainHalfCell & \resCodexPlainHalfCell & \resClaudePlainHalfCell & \resBfsPlainHalfCell & \resDfsPlainHalfCell & \resGreedyPlainHalfCell \\
At full budget & \resAbfPlainFullCell & \resCodexPlainFullCell & \resClaudePlainFullCell & \resBfsPlainFullCell & \resDfsPlainFullCell & \resGreedyPlainFullCell \\
\midrule
H2: movements claimed & \resAbfOpsClaimedCell & \resCodexOpsClaimedCell & \resClaudeOpsClaimedCell & \resBfsOpsClaimedCell & \resDfsOpsClaimedCell & \resGreedyOpsClaimedCell \\
H3: unreported transfers & \resAbfRepDef & \resCodexRepDef & \resClaudeRepDef & \resBfsRepDef & \resDfsRepDef & \resGreedyRepDef \\
\bottomrule
\end{tabular}

\end{table}

\paragraph{H3: Reliability.}
Agents can approach the tracers' reliability in assembling and
reporting evidence. The harness and Codex reach the fidelity of
\texttt{bfs}, and Claude Code falls below it (Table~\ref{tab:results}). Rejected exchanges include
contradictory amounts and receivers that the cited records leave
unbound to the exchange. Reporting also loses little: every agent omits at most
\resAbfRepDef\% of the transfers whose reference records it retrieved
(Table~\ref{tab:steps}), and the omissions may reflect overlooked or
misread records.

Agents may also cite evidence without retrieving it, a failure we call
\emph{overconfidence}. We found it in 29 of Claude Code's 31 runs,
despite the explicit instruction in every task to cite only served
records. Most of the unretrieved records it cites appear in the run
only as listing entries, and Claude Code may fill in their contents
from knowledge memorized during pre- or post-training. We impose the
instruction because forensic conclusions require rigor: language models
are known to hallucinate~\cite{huang2025hallucination}, and knowledge
left unchecked against the environment is prone to error. Under the
retrieval condition of edge recall (Section~\ref{sec:metrics}), Claude
Code reaches only \resClaudeRecall{}. Counting unretrieved citations
would raise it to \resClaudeRecallLenient{}, as the records it cites do
support most of its operations. Applying the same condition to fidelity
would lower Claude Code's to \resClaudeFidStrict{} and keep the harness
and Codex at \resAbfFidStrict{}. Apart from three such citations in 2 of
Codex's 31 runs, we observed this behavior only in Claude Code.

\section{Limitations}
\label{sec:discussion}
\label{sec:limits}

\paragraph{One incident.}
The suite covers one incident, rich in operations and ledgers
(Section~\ref{sec:case}). Further cases are in preparation to test how
far the results carry.

\paragraph{Training exposure.}
The incident is public and predates the evaluated models, which may
know it from pre- or post-training (Section~\ref{sec:steps}).

\paragraph{Conservative rules.}
The task rules halt where the records leave a continuation unresolved,
so edge recall measures recovery of a conservative subset of the
incident's flows.

\paragraph{Whole-system comparisons.}
Each comparison varies harness, model, and query policy together, so
its differences characterize whole systems. What agents would recover
with their unused budget remains unmeasured.

\section{Conclusion}
\label{sec:conclusion}

Public ledgers keep every movement of stolen funds on record, which
makes blockchain forensics a real-world investigative domain whose
conclusions can be checked mechanically: real adversaries supply its
difficulty, and the ledger keeps every claim verifiable. ABF-Bench
builds on this property. We formulate evidence-backed fund tracing as
an interactive, verifiable agent task; we construct the benchmark with
a reusable pipeline that turns a real laundering incident into
well-posed cross-chain tasks with conservative rules and query
budgets; and we propose the ABF agent harness, which outperforms both
general-purpose agents, including Codex on the same model.

The evaluation shows both the promise of agents and the gap that
remains: they acquire evidence faster than fixed-order tracers and
recognize custody that fixed rules walk past, yet they stop early and
end below breadth-first tracing. The harness's gain over Codex on the
same model shows that agent design moves these results, which leaves
room for better agents to close the gap.

Beyond measuring this gap, ABF-Bench opens blockchain forensics as a
verifiable task domain for language models. Every operation receives
a verdict checkable against the ledger, and real adversaries supply the
difficulty, which makes the benchmark a candidate environment for
training agents with verifiable rewards beyond code and mathematics.
In turn, bringing forensics into the scope of model training could let
advances in agents carry over to investigating laundering on public
ledgers. Further incidents will extend the suite.

\clearpage
\typeout{ABF-MAIN-PAGES: \number\numexpr\value{page}-1\relax}


\begin{thebibliography}{99}

\bibitem{meiklejohn2013fistful}
Sarah Meiklejohn, Marjori Pomarole, Grant Jordan, Kirill Levchenko, Damon
McCoy, Geoffrey M. Voelker, and Stefan Savage.
\newblock A Fistful of Bitcoins: Characterizing Payments Among Men with No
Names.
\newblock IMC, 2013.

\bibitem{pirolli2005sensemaking}
Peter Pirolli and Stuart Card.
\newblock The Sensemaking Process and Leverage Points for Analyst Technology
as Identified Through Cognitive Task Analysis.
\newblock International Conference on Intelligence Analysis, 2005.

\bibitem{weber2019aml}
Mark Weber, Giacomo Domeniconi, Jie Chen, Daniel Karl I. Weidele, Claudio
Bellei, Tom Robinson, and Charles E. Leiserson.
\newblock Anti-Money Laundering in Bitcoin: Experimenting with Graph
Convolutional Networks for Financial Forensics.
\newblock KDD Workshop on Anomaly Detection in Finance, 2019.

\bibitem{elliptic2}
Claudio Bellei, Muhua Xu, Ross Phillips, Tom Robinson, Mark Weber, Tim Kaler,
Charles E. Leiserson, Arvind, and Jie Chen.
\newblock The Shape of Money Laundering: Subgraph Representation Learning on
the Blockchain with the Elliptic2 Dataset.
\newblock KDD Workshop on Machine Learning in Finance, 2024. arXiv:2404.19109.

\bibitem{tracer}
Zhiying Wu, Jieli Liu, Jiajing Wu, Zibin Zheng, and Ting Chen.
\newblock TRacer: Scalable Graph-Based Transaction Tracing for Account-Based
Blockchain Trading Systems.
\newblock \emph{IEEE Transactions on Information Forensics and Security},
18:2609--2621, 2023.

\bibitem{connector}
Dan Lin, Jiajing Wu, Yuxin Su, Ziye Zheng, Yuhong Nan, Qinnan Zhang, Bowen
Song, and Zibin Zheng.
\newblock Connector: Enhancing the Traceability of Decentralized Bridge
Applications via Automatic Cross-Chain Transaction Association.
\newblock \emph{IEEE Transactions on Information Forensics and Security},
20:7588--7601, 2025.

\bibitem{abctracer}
Dan Lin, Ziye Zheng, Jiajing Wu, Jingjing Yang, Kaixin Lin, Huan Xiao, Bowen
Song, and Zibin Zheng.
\newblock Track and Trace: Automatically Uncovering Cross-Chain Transactions in
the Multi-Blockchain Ecosystems.
\newblock \emph{IEEE Transactions on Services Computing}, 18:4291--4303, 2025.

\bibitem{faultseeker}
Kairan Sun, Zhengzi Xu, Kaixuan Li, Lyuye Zhang, Yuqiang Sun, Liwei Tan, and
Yang Liu.
\newblock FaultSeeker: LLM-Empowered Framework for Blockchain Transaction Fault
Localization.
\newblock In \emph{IEEE/ACM ASE}, pages 104--116, 2025.

\bibitem{txray}
Ziyue Wang, Jiangshan Yu, Kaihua Qin, Dawn Song, Arthur Gervais, and Liyi Zhou.
\newblock TxRay: Agentic Postmortem of Live Blockchain Attacks.
\newblock arXiv:2602.01317, 2026.

\bibitem{locard}
Xiaohang Yu and William J. Knottenbelt.
\newblock LOCARD: An Agentic Framework for Blockchain Forensics.
\newblock arXiv:2604.04211, 2026.

\bibitem{txsum}
Zifan Peng, Jingyi Zheng, Yule Liu, Huaiyu Jia, Qiming Ye, Jingyu Liu, Xufeng
Yang, Mingchen Li, Qingyuan Gong, Xuechao Wang, and Xinlei He.
\newblock TxSum: User-Centered Ethereum Transaction Understanding with
Micro-Level Semantic Grounding.
\newblock EMNLP, 2026. arXiv:2512.06933.

\bibitem{risktagger}
Dan Lin, Yanli Ding, Weipeng Zou, Jiajing Wu, Zhiyin Wu, Jiachi Chen, Xiapu
Luo, and Zibin Zheng.
\newblock RiskTagger: Evidence-Guided LLM Agent for Post-Incident Forensic
Analysis of Money Laundering in Web3.
\newblock arXiv:2510.17848v2, 2026.

\bibitem{cryptobench}
Jiacheng Guo, Suozhi Huang, Zixin Yao, et al.
\newblock CryptoBench: A Dynamic Benchmark for Expert-Level Evaluation of LLM
Agents in Cryptocurrency.
\newblock arXiv:2512.00417, 2025.

\bibitem{aiabc}
Gate.
\newblock AI-ABC: Agent Benchmark for Crypto.
\newblock \url{https://github.com/gate/ai-abc}, 2026.

\bibitem{amlguard}
Hao Wu, Haijun Wang, Shangwang Li, Yin Wu, Ming Fan, Ting Liu, and Xiapu Luo.
\newblock Tracing the Shadows: Automatic Tracking and Analysis of Crypto Money
Laundering via Transaction Semantic Analysis.
\newblock arXiv:2607.18869, 2026.

\bibitem{xsplicer}
Dan Lin, Huan Xiao, Ziwei Li, Xiapu Luo, Jiachi Chen, Jiajing Wu, and Zibin
Zheng.
\newblock Bridging the Opacity: Evidence-Backed Cross-Chain Transaction
Correspondence Reconstruction Across Heterogeneous Blockchains.
\newblock arXiv:2609.18158, 2026.

\bibitem{swebench}
Carlos E. Jimenez, John Yang, Alexander Wettig, Shunyu Yao, Kexin Pei,
Ofir Press, and Karthik Narasimhan.
\newblock SWE-bench: Can Language Models Resolve Real-World GitHub Issues?
\newblock ICLR, 2024.

\bibitem{deepseekr1}
DeepSeek-AI.
\newblock DeepSeek-R1: Incentivizing Reasoning Capability in LLMs via
Reinforcement Learning.
\newblock arXiv:2501.12948, 2025.

\bibitem{hu2025memory}
Yuyang Hu, Shichun Liu, Yanwei Yue, et al.
\newblock Memory in the Age of AI Agents.
\newblock arXiv:2512.13564, 2025.

\bibitem{zhang2026landscape}
Guibin Zhang, Hejia Geng, Xiaohang Yu, et al.
\newblock The Landscape of Agentic Reinforcement Learning for LLMs: A Survey.
\newblock \emph{Transactions on Machine Learning Research}, 2026.

\bibitem{brown2020gpt3}
Tom B. Brown, Benjamin Mann, Nick Ryder, Melanie Subbiah, et al.
\newblock Language Models are Few-Shot Learners.
\newblock NeurIPS, 2020.

\bibitem{huang2025hallucination}
Lei Huang, Weijiang Yu, Weitao Ma, Weihong Zhong, Zhangyin Feng, Haotian
Wang, et~al.
\newblock A Survey on Hallucination in Large Language Models: Principles,
Taxonomy, Challenges, and Open Questions.
\newblock \emph{ACM Transactions on Information Systems}, 43(2):1--55, 2025.

\bibitem{hle}
Long Phan, Alice Gatti, Ziwen Han, et al.
\newblock Humanity's Last Exam.
\newblock arXiv:2501.14249, 2025.

\bibitem{frontiermath}
Elliot Glazer, Ege Erdil, Tamay Besiroglu, et al.
\newblock FrontierMath: A Benchmark for Evaluating Advanced Mathematical
Reasoning in AI.
\newblock arXiv:2411.04872, 2024.

\bibitem{fbi2025psa}
Federal Bureau of Investigation, Internet Crime Complaint Center.
\newblock Public Service Announcement I-022625-PSA: North Korea Responsible
for \$1.5 Billion Bybit Hack.
\newblock \url{https://www.ic3.gov/PSA/2025/PSA250226}, 2025.

\bibitem{hackscan2025}
Bybit.
\newblock HackScan public blacklist: Bybit incident address set.
\newblock \url{https://hackscan.hackbounty.io/public/hack-address.json},
2025.

\bibitem{moser2014taint}
Malte M\"oser, Rainer B\"ohme, and Dominic Breuker.
\newblock Towards Risk Scoring of Bitcoin Transactions.
\newblock In \emph{Financial Cryptography and Data Security Workshops}, pages
16--32, 2014.

\bibitem{tironsakkul2019taint}
Tin Tironsakkul, Manuel Maarek, Andrea Eross, and Mike Just.
\newblock Probing the Mystery of Cryptocurrency Theft: An Investigation into
Methods for Taint Analysis.
\newblock arXiv:1906.05754, 2019.

\bibitem{chainalysis2024}
Chainalysis.
\newblock \$2.2 Billion Stolen in Crypto in 2024 but Hacked Volumes Stagnate.
\newblock Chainalysis blog, 19 December 2024.
\url{https://www.chainalysis.com/blog/crypto-hacking-stolen-funds-2025/}.

\bibitem{tsuchiya2025poisoning}
Taro Tsuchiya, Jin-Dong Dong, Kyle Soska, and Nicolas Christin.
\newblock Blockchain Address Poisoning.
\newblock USENIX Security, pp.~1243--1262, 2025.

\bibitem{wang2023mixers}
Zhipeng Wang, Stefanos Chaliasos, Kaihua Qin, Liyi Zhou, Lifeng Gao,
Pascal Berrang, Benjamin Livshits, and Arthur Gervais.
\newblock On How Zero-Knowledge Proof Blockchain Mixers Improve, and Worsen
User Privacy.
\newblock ACM Web Conference, pp.~2022--2032, 2023.

\bibitem{webarena}
Shuyan Zhou, Frank F. Xu, Hao Zhu, Xuhui Zhou, Robert Lo, Abishek Sridhar,
et~al.
\newblock WebArena: A Realistic Web Environment for Building Autonomous
Agents.
\newblock ICLR, 2024. arXiv:2307.13854.

\bibitem{taubench}
Shunyu Yao, Noah Shinn, Pedram Razavi, and Karthik Narasimhan.
\newblock $\tau$-bench: A Benchmark for Tool-Agent-User Interaction in
Real-World Domains.
\newblock arXiv:2406.12045, 2024.

\bibitem{cybench}
Andy K. Zhang, Neil Perry, Riya Dulepet, Joey Ji, Celeste Menders,
Justin W. Lin, et~al.
\newblock Cybench: A Framework for Evaluating Cybersecurity Capabilities and
Risks of Language Models.
\newblock ICLR, 2025. arXiv:2408.08926.

\bibitem{browsecomp}
Jason Wei, Zhiqing Sun, Spencer Papay, Scott McKinney, Jeffrey Han, Isa
Fulford, et~al.
\newblock BrowseComp: A Simple Yet Challenging Benchmark for Browsing Agents.
\newblock arXiv:2504.12516, 2025.

\bibitem{zhu2025abc}
Yuxuan Zhu, Tengjun Jin, Yada Pruksachatkun, Andy Zhang, Shu Liu, Sasha
Cui, Sayash Kapoor, Shayne Longpre, Kevin Meng, Rebecca Weiss, Fazl
Barez, Rahul Gupta, Jwala Dhamala, Jacob Merizian, Mario Giulianelli,
Harry Coppock, Cozmin Ududec, Jasjeet Sekhon, Jacob Steinhardt, Antony
Kellermann, Sarah Schwettmann, Matei Zaharia, Ion Stoica, Percy Liang,
and Daniel Kang.
\newblock Establishing Best Practices for Building Rigorous Agentic
Benchmarks.
\newblock arXiv:2507.02825, 2025.

\end{thebibliography}
\end{document}